\documentclass[%
 reprint,
superscriptaddress,
 amsmath,amssymb,
 aps,
]{revtex4-2}

\usepackage{amsthm}
\theoremstyle{definition}

\usepackage[format=plain]{caption}
\usepackage{braket}

\usepackage{graphicx}
\usepackage{dcolumn}
\usepackage{bm}
\usepackage{xcolor}

\usepackage{algpseudocode}
\usepackage{algorithm}

\usepackage{indentfirst}
\usepackage{hyphenat}
\usepackage{xurl}
\usepackage{hyperref}
\usepackage[normalem]{ulem}
\usepackage{ragged2e}

\usepackage{lipsum}

\expandafter\def\expandafter\normalsize\expandafter{%
    \normalsize%
    \setlength\abovedisplayskip{3pt}%
    \setlength\belowdisplayskip{4pt}%
    \setlength\abovedisplayshortskip{-1pt}%
    \setlength\belowdisplayshortskip{4pt}%
}

\begin{document}

\preprint{APS/123-QED}

\title{Healing Regimes for Microscopic Wounds in the Vertex Model of Cell Tissues}

\author{R. F. Almada}
\email{rffalmada@fc.ul.pt}
\affiliation{Centro de Física Teórica e Computacional, Faculdade de Ciências da Universidade de Lisboa, 1749-016 Lisboa, Portugal}
\affiliation{Departamento de Física, Faculdade de Ciências da Universidade de Lisboa, 1749-016 Lisboa, Portugal}
\author{N. A. M. Araújo}
\email{nmaraujo@fc.ul.pt}
\affiliation{Centro de Física Teórica e Computacional, Faculdade de Ciências da Universidade de Lisboa, 1749-016 Lisboa, Portugal}
\affiliation{Departamento de Física, Faculdade de Ciências da Universidade de Lisboa, 1749-016 Lisboa, Portugal}

\author{P. Patrício}
\email{pedro.patricio@isel.pt}
\affiliation{Centro de Física Teórica e Computacional, Faculdade de Ciências da Universidade de Lisboa, 1749-016 Lisboa, Portugal}
\affiliation{Instituto Superior de Engenharia de Lisboa, Instituto Politécnico de Lisboa, 1959-007 Lisboa, Portugal}

\date{\today}

\begin{abstract}

Wounds in epithelial tissues compromise their vital role in homeostasis. A rapid and efficient wound healing encompasses different mechanisms, which includes the formation of a contractile actin-myosin cable around its edge, known as the purse-string mechanism. We combine mean-field calculations and numerical simulations of the Vertex model to study the interplay between tissue properties and the purse-string mechanism and its impact on the healing process. We find different regimes, where the wound opens, closes partially or completely. We also derive an analytic expression for the closure time which is validated by numerical simulations. 
This study establishes under which conditions the purse-string mechanism suffices for closure, providing an analytical mean-field expression for the respective thresholds. 

\end{abstract}

\maketitle

\section{Introduction}

Epithelial tissues are groups of cells acting collectively to cover the body and organs, maintaining homeostasis in living systems \cite{lodish,lecuit2007}.
The integrity of these tissues is compromised by the formation of wounds. 
They are discontinuities or gaps in the tissue, and to restore tissue integrity, specific mechanisms come into play, through the process of wound healing \cite{Begnaud2016}.

The primary closure process involves a combination of the purse-string and cell crawling mechanisms \cite{nodder1997, tamada2007two, klarlund2012dual}. 
The former consists of the formation of a purse-string-like structure of  F-actin and  myosin filaments encircling the wound, which dynamically contracts, with the filaments anchored by adherens and tight junctions at the edge of the wound. 
This mechanism is also critical in cellular extrusion, morphogenesis \cite{martin1992, danjo1998,bement1999wound, kiehart1999wound, jacinto2002, sonnemann2011, villars2022}, and cell migratory processes \cite{mitchison1996, jacinto2000, wood2002, martin2004,anon2012cell,theveneau2013}.  
Cell crawling is characterized by cells extending protrusions called lamellipodia at the edge of the wound, often accompanied by a polarity switch.

A body of experimental studies aided by numerical simulations reveals that the dynamics of closure is affected by various properties of the tissue and surrounding medium, such as the elasticity of the epithelial sheet and the substrate rigidity \cite{arciero2011}, wound shape \cite{arciero2013using, ravasio2015}, tissue fluidity \cite{staddon2018, tetley2019}, focal adhesions \cite{brugues2014}, substrate \cite{vedula2015, ravasio2015regulation} and cell adhesiveness \cite{noppe2015, carvalho2018}, and actin polymerization and polarity \cite{chen2019large, wei2020actin}. In the study of tissue mechanics, different models are employed, depending on the level of description, from detailed cell-based models to continuous, coarse-grained descriptions, which access the collective dynamics of the tissue, usually occurring in longer length and time scales \cite{camley2017, salbreux2017, hirashima2017, saw2018, alert2020}. 
While the cellular dynamics involve a range of complex biochemical phenomena, experimental and theoretical results suggest that a mechanical framework captures a wide range of behaviors of the tissue \cite{camley2017,alert2020}.
Vertex models are a notable example due to their simplicity and ability to link physical properties of tissues with geometric and topological properties of the network of cells \cite{alert2020, gomezgalvez2021}. 
Thus, these models have succeeded in describing static and dynamic properties of cell tissues \cite{manning2016, staddon2018, ajeti2019, tetley2019, manning2021,atia2021cell}.
We aim to use the vertex model to understand how the tissue properties correlate with cell parameters and, in particular, their relation with the purse-string mechanism, driving the closure process. 

This paper is organized in the following way.
In section \ref{sec:2}, we describe the Vertex model and derive a mean-field perturbative analysis to identify the different healing regimes.
In section \ref{sec:3},  we discuss numerical results based on a finite gradient implementation of the Vertex model for different sets of parameters to evaluate the role of spatial heterogeneities not captured by the mean-field description. 
In section \ref{sec:4}, we discuss the results and draw comparisons with other results from the literature. 

\section{Wound closure with network rearrangement} 
\label{sec:2}

\color{black}
We first describe the Vertex model and its dynamics for an isotropic tissue. We then perform mean-field calculations to determine how the asymptotic value of the wound perimeter depends on the different model parameters, identifying three healing regimes.
\color{black}
\subsection{Vertex Model}
\begin{figure}
    \centering
    \includegraphics[width=0.45\textwidth]{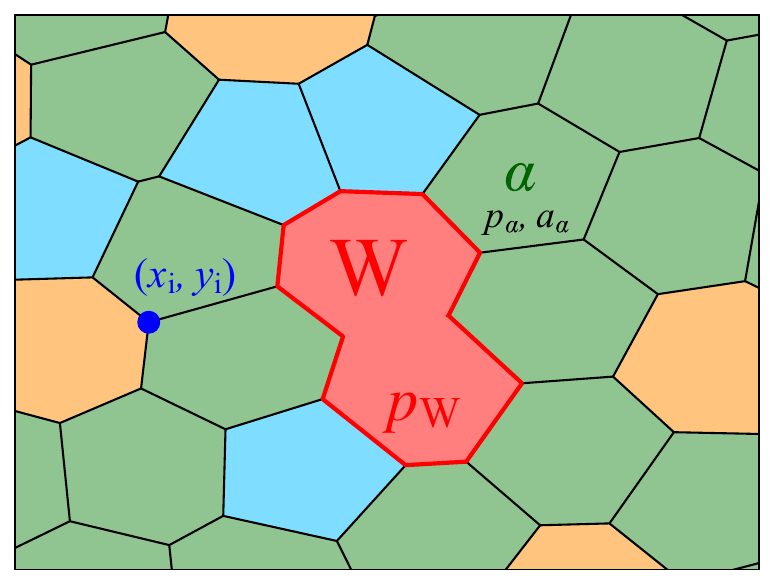}
    \caption{\small A sketch of the cell network, where $(x_i, y_i)$ corresponds the coordinates of vertex $i$, highlighted in dark blue,
    $\alpha$ correspond to a polygonal cell, described by a set of vertices and parametrized by its perimeter $p_\alpha$ and area $a_\alpha$. In red is the wound $\mathrm{W}$ of perimeter $p_\mathrm{W}$. The other cells are colored based on the number of edges, with pentagons in light blue, hexagons in green and larger polygons in orange.}
    \label{fig:fig1}  
\end{figure}

\color{blue}
\color{black}
A 2D tissue of $N_\mathrm{C}$ polygonal cells is represented by vertices $i = 1, \text{...}, N_\mathrm{V}$  with coordinates $ \mathbf{r}_i = (x_i, y_i)$, as shown in Fig. (\ref{fig:fig1}). The set of all the coordinates is $\mathbf{R} = \{\mathbf{r}_1,..., \mathbf{r}_{N_\mathrm{V}}\}$. 
A cell $\alpha$ is the polygon defined by an ordered set of $m$ vertices $ \{i_{\alpha_1}, ..., i_{\alpha_m}\}$. Consecutive vertices are adjacent to each other, and the last vertex $i_{\alpha_m}$ is adjacent to the first. 
Under the assumption that all cells are identical, the total free energy $\mathcal{F}$ is defined as:
\begin{equation}
\begin{split}
\small
\mathcal{F}(\mathbf{R}) = \frac{1}{2}\sum_{\alpha = 1}^{N_\mathrm{C}}{\left[K\left(A_\alpha-A_\mathrm{0}\right)^2+ \Gamma\left(P_\alpha-P_\mathrm{0}\right)^2\right]} +  \Lambda_\mathrm{W} P_\mathrm{W},
 \label{eqn:eq1}
 \end{split}
\end{equation}
where $A_\alpha$ and $A_\mathrm{0}$ are the actual and  preferred areas of cell $\alpha$ and $K$ is the area stiffness due to the incompressibility of the cell in 3D. $P_\alpha$ and $P_\mathrm{0}$ are the actual and preferred perimeters and $\Gamma$ corresponds to perimeter stiffness due to cell contractility. The sum is over all cells in the tissue. $\Lambda_\mathrm{W}$ is the  effective wound tension due to the balance between adhesive line tension and purse-string contraction, and $P_\mathrm{W}$ is the wound perimeter (see suppl. section A). 

We neglect inertial effects and thus assume an overdamped regime. The $N_\mathrm{V}$-coupled equations of motion are:
\begin{equation}
     \mu \frac{d{\mathbf{r}}_i}{d t} = -\frac{\delta\mathcal{F}}{\delta{\mathbf{r}}_i},
     \label{eqn:eq2}
\end{equation}
where $\mu$ is the friction coefficient.

We define a dimensionless free energy $\Tilde{\mathcal{F}} = {\mathcal{F}}/({K A_\mathrm{0}^2})$, which, from Eq. (\ref{eqn:eq1}), gives:
\begin{equation}
\begin{split}
 \Tilde{\mathcal{F}}(\mathbf{r}) = \sum_{\alpha = 1}^{N_\mathrm{C}}{\left[\frac{1}{2}(a_\alpha-1)^2+ \frac{\gamma}{2}(p_\alpha-p_\mathrm{0})^2\right]} +  \lambda_\mathrm{W} p_\mathrm{W},
 \normalsize
 \label{eqn:eq3}
 \end{split}
\end{equation}
where $a_\alpha  = A_\alpha/A_\mathrm{0}$,  $p_\alpha  = P_\alpha/\sqrt{A_\mathrm{0}}$, and $p_\mathrm{W}  = P_\mathrm{W}/\sqrt{A_\mathrm{0}}$ are the dimensionless area and perimeter of a cell, and perimeter of the wound, respectively.  $\gamma = \Gamma/(K A_\mathrm{0})$ is the dimensionless ratio of perimeter to area stiffness, $p_\mathrm{0}  = P_\mathrm{0}/\sqrt{A_\mathrm{0}}$ the shape parameter, and $\lambda_\mathrm{W} = \Lambda_\mathrm{W}/(KA_\mathrm{0}^{3/2})$ the dimensionless effective wound tension. 
The $N_\mathrm{V}$-coupled equations of motion are also redimensionalized as:
\begin{equation}
     \frac{d \Tilde{\mathbf{r}}_i}{d \xi} = -\frac{\delta\Tilde{\mathcal{F}}}{\delta\Tilde{\mathbf{r}}_i},
     \label{eqn:eq4}
\end{equation}
where $\xi = t/\tau_\mu$ and $\Tilde{\mathbf{r}}_i = \mathbf{r}_i/\sqrt{A_\mathrm{0}}$ are the dimensionless time and space, respectively, ${\delta\Tilde{\mathcal{F}}}/{\delta\Tilde{\mathbf{r}}_i}$ is the derivative of $\Tilde{\mathcal{F}}$ with respect to $\Tilde{\mathbf{r}}_i$, and $\tau_\mu = \mu/K A_\mathrm{0}$ is the relaxation time scale associated with the friction coefficient $\mu$.

This is a simplified model, as it assumes a constant purse-string coefficient. A previous work considered an higher-order dependency on the wound perimeter and, while no explicit reasoning for this choice is provided, we suspect it is to ensure a symmetric energy term \cite{noppe2015}.
We consider a linear dependency in the wound perimeter, rather than a quadratic dependency, as in previous works, for example, Ref.\cite{tetley2019}.

\color{black}

\subsection{Mean-field calculation}

\label{appendix:meanfieldwork}
Let us assume a regular network of vertices defining $N_\mathrm{C}$ cells of equal shape and size ($a_\alpha = a, p_\alpha = p$). The tissue is initially at equilibrium $(a(0) = 1)$.
For a wound of initial area $a_\mathrm{W}(0) = N_\mathrm{W}a(0)$, where $N_\mathrm{W}\ll N_\mathrm{C}$ is the number of removed cells, the Eq. (\ref{eqn:eq3}) simplifies as:
\begin{small}
\begin{equation}
    \begin{split}
        \Tilde{\mathcal{F}} = (N_\mathrm{C}-N_\mathrm{W})\left[\frac{1}{2}(a-1)^2+ \frac{\gamma}{2}(p-p_\mathrm{0})^2\right] + \lambda_\mathrm{W} p_\mathrm{W}.
    \end{split}
    \label{eqn:eq5}
\end{equation}
\end{small}

For a fixed external boundary, the total area of the tissue is constant, $(N_\mathrm{C}-N_\mathrm{W})a + a_\mathrm{W} = N_\mathrm{C}a(0)$.
We define the cell isoperimetric ratio as $\rho = p^2/a$.
Plugging this relation into Eq. (\ref{eqn:eq5}), we obtain:
\begin{equation}
\begin{split}
      \Tilde{\mathcal{F}}(p_\mathrm{W}) = \frac{1}{2(N_\mathrm{C}-N_\mathrm{W})}\left(N_\mathrm{W}-\rho_\mathrm{W} p_\mathrm{W}^2\right)^2 + \\ \frac{\gamma}{2\rho}\left(\sqrt{N_\mathrm{C}-\rho_\mathrm{W} p_\mathrm{W}^2}-p_\mathrm{0}\rho^{\frac{1}{2}}\sqrt{N_\mathrm{C}-N_\mathrm{W}}\right)^2 + \lambda_\mathrm{W} p_\mathrm{W},
\end{split}     
\label{eqn:eq6}
\end{equation}
where we replaced $a_\mathrm{W} = p_\mathrm{W} ^ 2 \rho_\mathrm{W}$, defining $\rho_\mathrm{W}$ as the wound isoperimetric ratio.  
In the overdamped regime, $p_\mathrm{W}$ changes very slowly. Following from the Onsager principle, the time evolution of the wound size is determined by the relation:

\begin{equation}
    \tau_\mathrm{p} \frac{d p_\mathrm{W}}{d t} = - \frac{\partial \Tilde{\mathcal{F}}}{\partial p_\mathrm{W}},
    \label{eqn:eq7}
\end{equation}
where $\tau_\mathrm{p} = \tau_\mu\phi_\mathrm{W}$ is a timescale of the wound frictional response \cite{wang2021, binder1973, santambrogio2017}, with $\phi_\mathrm{W}$ being a parameter dependent on wound shape (see suppl. section B).
The derivative ${\partial \Tilde{\mathcal{F}}}/{\partial p_\mathrm{W}}$ is given by the following expression:
\begin{equation}
\begin{split}
   \frac{\partial \Tilde{\mathcal{F}}}{\partial p_\mathrm{W}} = \gamma\frac{ \rho_\mathrm{W}}{\rho}\left( p_\mathrm{0} \rho^\frac{1}{2}\sqrt{\frac{ N_\mathrm{C} - N_\mathrm{W}}{N_\mathrm{C} - p_\mathrm{W}^2 \rho_\mathrm{W}}} - 1\right)p_\mathrm{W} \\ + \frac{2}{N_\mathrm{C}-N_\mathrm{W}}\rho_\mathrm{W}\left(\rho_\mathrm{W}p_\mathrm{W}^2- N_\mathrm{W}\right)p_\mathrm{W}  + \lambda_\mathrm{W} = -\frac{d p_\mathrm{W}}{d \tau},
\end{split}
\label{eqn:eq8}
\end{equation}
where $\tau_\mathrm{p} {d p_\mathrm{W}}/{d t}$ is rewritten through a change of variables of the form $\tau = t /\tau_\mathrm{p}$. Since $N_\mathrm{W} \ll N_\mathrm{C}$, we Taylor expand in $ \varepsilon = {1}/{N_\mathrm{C}}$, and solve using perturbative methods \cite{nayfeh2011introduction, guckenheimer2013}.

\label{sec:qualanal}

Even in the absence of an effective wound tension ($\lambda_\mathrm{W} = 0$), there are three stationary solutions for $p_\mathrm{W}$: $p_\mathrm{W} = 0$ and $p_\mathrm{W}\sim \pm \sqrt{p^\text{bif}_\mathrm{0}- p_\mathrm{0}}$.
Since $p_\text{W} < 0$ has no physical meaning, only two are relevant: $p_\text{W} = 0$ corresponds to a closed wound and $p_\text{W} > 0$ to one that does not close. $p^\text{bif}_\mathrm{0}$, defined as:
\begin{equation}
    p^\text{bif}_\mathrm{0} \approx \rho^{-\frac{1}{2}}\left[1 + \left(\frac{1}{2}+2\frac{\rho}{\gamma}\right)N_\mathrm{W}\varepsilon\right],
    \label{eqn:eq9}
\end{equation}
is the threshold value for $p_\mathrm{0}$ between these two limits (for $\lambda_\mathrm{W} = 0$), determined to first order in $\varepsilon$.  As $\varepsilon \rightarrow 0$, $p^\text{bif}_\mathrm{0}$ converges to $\rho^{-\frac{1}{2}}$, coinciding with the critical point for tissue glass transition from a solid-like to fluid-like state, a result previously reported in \cite{farhadifar2007, staple2010}. The link between the critical point and wound closure has been established previously \cite{noppe2015,tetley2019}. Given this correspondence with the critical point for the tissue transition, we define $p^\text{crit}_\mathrm{0} = \rho^{-\frac{1}{2}}$. For reference, the value for a pentagon is $3.812$ and for an hexagon it is $3.722$.
\begin{figure}
    \centering
    \includegraphics[width=0.475\textwidth]{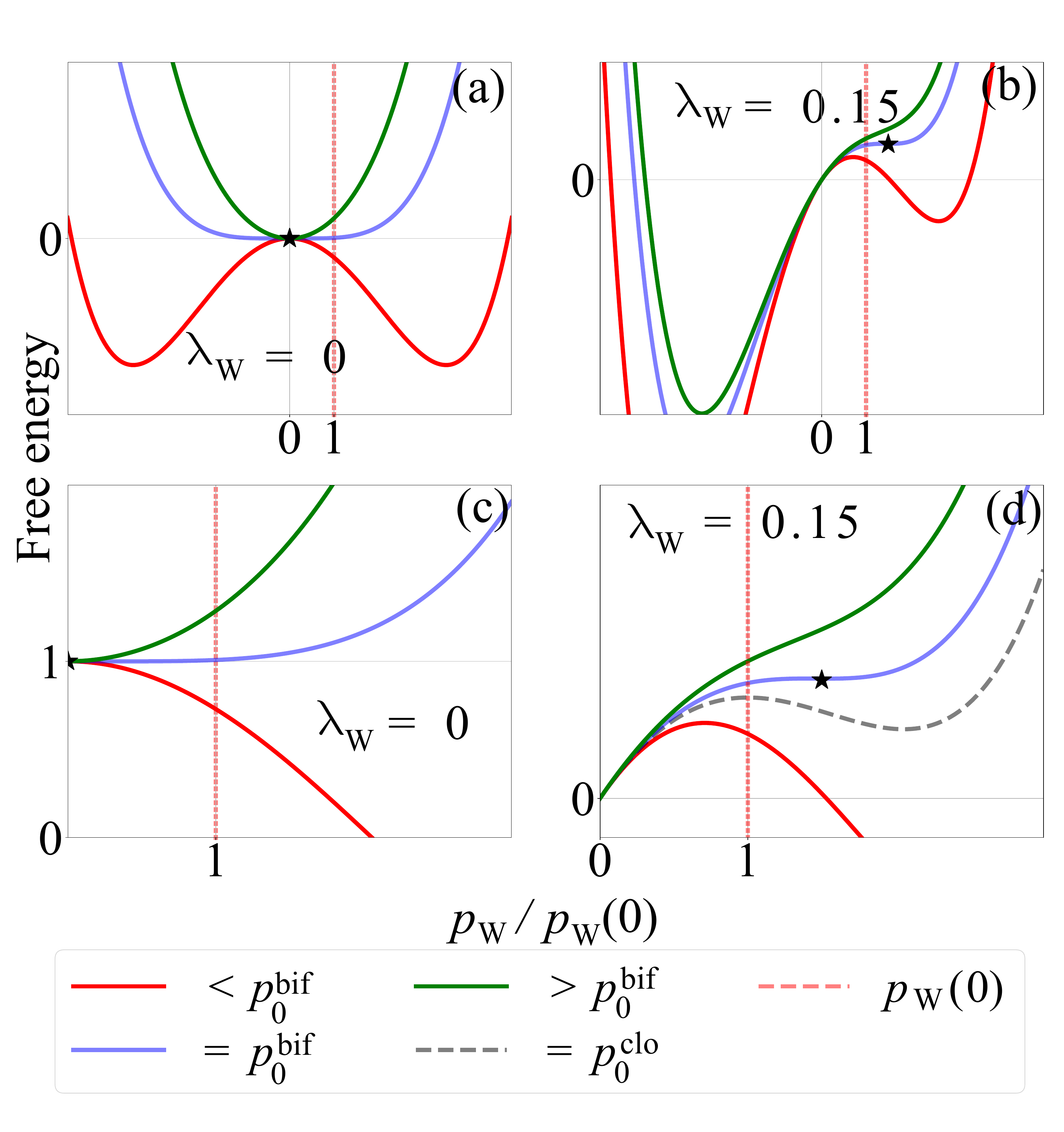}
    \caption{\footnotesize (a,b) - Free energy as function of wound perimeter (normalized to the initial perimeter $\sqrt{N_\mathrm{W}/\rho_\mathrm{W}}$) in the absence (a) and presence (b) of effective purse tension for different cell shape parameters ($p_\mathrm{0}$) relative to the critical threshold ($p^\text{bif}_\mathrm{0}$). The bifurcation point marks the shift from two energy minima and a local energy maximum to one energy minimum. (c,d) - Free energy zoomed in the neighborhood of the initial perimeter. The dashed black curve indicates the point at which $p_\mathrm{W}(0)$ is the local energy maximum. The parameters used to compute these images correspond to $N_\mathrm{C} = 100, N_\mathrm{W} = 1, \gamma = 1, \rho= \rho_W$ for an hexagonal network.}
    \label{fig:fig2}
\end{figure}

For $p_\mathrm{0} < p^\text{bif}_\mathrm{0}$, $\Tilde{\mathcal{F}}$ exhibits two minima which are stable fixed points and one unstable fixed point for $p_\mathrm{W} = 0$. The wound never closes (Figs. \ref{fig:fig2}.a and \ref{fig:fig2}.c, red curve). By contrast, for $p_\mathrm{0} \geq p^\text{bif}_\mathrm{0}$, $\Tilde{\mathcal{F}}$ has only one minimum at $p_\mathrm{W} = 0$, corresponding to wound closure (Figs. \ref{fig:fig2}.a and \ref{fig:fig2}.c, green curve).
At $p_\mathrm{0} = p^\text{bif}_\mathrm{0}$, this minimum bifurcates (Figs. \ref{fig:fig2}.a and \ref{fig:fig2}.c, blue curve) at $\lambda_\mathrm{W} = 0$, which is a property of supercritical pitchfork bifurcations \cite{itzykson_drouffe_1989, guckenheimer2013}. 

When $\lambda_\mathrm{W} \neq 0$, $\Tilde{\mathcal{F}}$ is no longer symmetric (Fig. \ref{fig:fig2}.b, red curve), and the maximum shifts for positive values of $p_\mathrm{W}$. Whether the wound closes or not depends on the initial wound perimeter $p_\mathrm{W}(0) = \sqrt{N_\mathrm{W}/\rho_\mathrm{W}}$. When the wound is smaller than the local energy maximum, $\partial \Tilde{\mathcal{F}}/\partial p_\mathrm{W}$ is positive, then wound closure occurs (Fig. \ref{fig:fig2}.d, dashed gray, blue and green curves); for negative $\partial \Tilde{\mathcal{F}}/\partial p_\mathrm{W}$ the wound never closes (Fig. \ref{fig:fig2}.d, red curve). The dashed curve indicates the point at which $p_\mathrm{W}(0)$ is the local maximum of the energy.

For a given initial wound size $p_\mathrm{W}(0)$ and shape parameter $p_\mathrm{0}$, a minimum effective wound tension $\lambda^\text{clo}_\mathrm{W}$ is necessary for closure. At this point, the free energy derivative at $p_\mathrm{W}(0)$ becomes null.
The transition between wound opening and wound closure occurs for: 

\begin{equation}
    \lambda^\text{clo}_\mathrm{W} = \gamma \sqrt{N_\mathrm{w}} \left(\frac{\rho_\mathrm{W}}{\rho}\right)^{1/2} (p^\text{crit}_\mathrm{0}-p_\mathrm{0}).
    \label{eqn:eq10}
\end{equation}
Note that $\lambda_\mathrm{W}$ accounts for purse-string contraction and neighboring adhesive tensions (see suppl. section A).
Equation (\ref{eqn:eq10}) is a mean-field exact result, determined without any perturbative methods. 

For each $\lambda_\mathrm{W}$, as we change $p_\mathrm{0}$, we also observe a bifurcation above a threshold. 
At the bifurcation point, the local energy maximum and minimum merge into a degenerate point (Fig. \ref{fig:fig2}.b, blue curve), which also corresponds to a pitchfork with broken symmetry. The condition for bifurcation is given by ${\partial^2 \Tilde{\mathcal{F}}}/{\partial p_\mathrm{W}^2} = {\partial \Tilde{\mathcal{F}}}/{\partial p_\mathrm{W}} = 0$. Beyond this point, only one minimum at a negative wound perimeter exists (Fig. \ref{fig:fig2}.b, green curve). Given that the derivative is positive and the only existing minimum is negative, all closure processes are successful.  

The bifurcation threshold can not be determined exactly in the mean-field approach. However, for $\lambda_\mathrm{W}$ of order $\varepsilon$, we may expect that the bifurcation threshold occurs at $p_\mathrm{0}$ near $p^\text{bif}_\mathrm{0}$ (see Eq. (\ref{eqn:eq9})). Assuming  ($p_\mathrm{0} - p^\text{bif}_\mathrm{0} \sim \mathcal{O}(\varepsilon)$), a perturbative analysis gives:
\begin{equation}
\begin{split}
        \lambda_\mathrm{W} = \lambda^\text{bif}_\mathrm{W} \approx  N_\mathrm{C}^{\frac{1}{2}} \frac{\rho_\mathrm{W}}{2} (2+\frac{\gamma}{2\rho})^{-1} (3\gamma \rho^{-1/2})^\frac{3}{2}(p^\text{bif}_\mathrm{0}-p_\mathrm{0})^\frac{3}{2}.
\end{split}
    \label{eqn:eq11}
\end{equation}
This bifurcation threshold is valid only to first order in $\varepsilon$, and so it has an explicit dependency on the total number of cells in the tissue.

\begin{figure}
    \centering
    \includegraphics[width=0.475\textwidth]{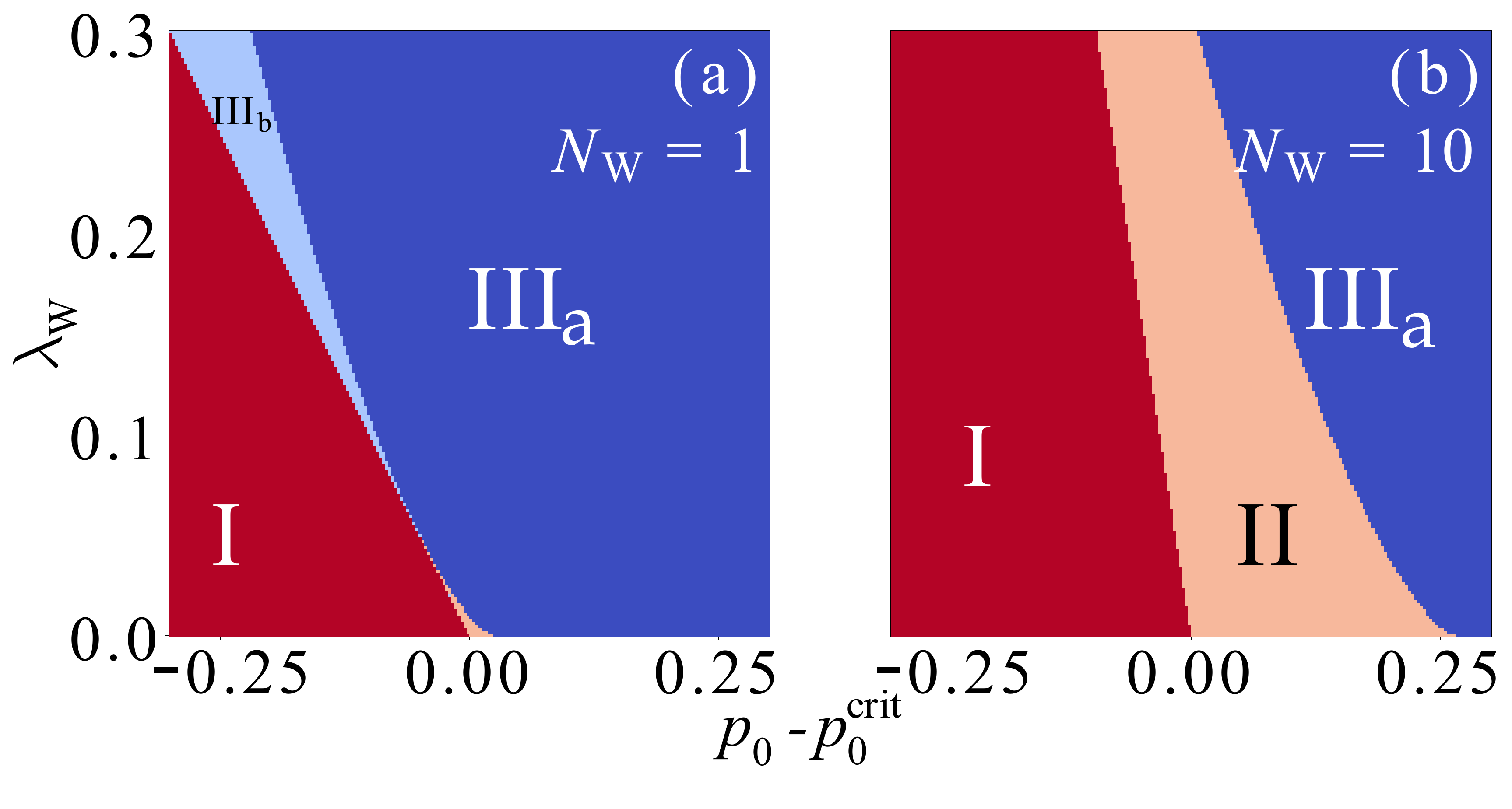}
    \caption{\footnotesize Two-parameter diagram for wound closure with an initial perimeter corresponding a) to $N_\mathrm{W} = 1$ ($\varepsilon = 0.01$); b) $N_\mathrm{W} = 10$ ($\varepsilon = 0.1$). With $\gamma = 1$, $\rho=\rho_\mathrm{W}$ for an hexagonal network.}
    \label{fig:fig3}
\end{figure}

$\lambda_\mathrm{W}$ and $p_\mathrm{0}$ define a two-parameter space which is divided into three regimes (Fig. \ref{fig:fig3}.a):

\begin{itemize}
    \item Regime I - the wound opens - the free energy has two minima and $\partial \Tilde{\mathcal{F}}/\partial p_\mathrm{W} < 0$, so the wound perimeter converges to a positive value greater than $p_\mathrm{W}(0)$ (see Fig 1.a of suppl. section C);
    \item Regime II - the wound closes partially - the free energy has two minima, but $\partial \Tilde{\mathcal{F}}/\partial p_\mathrm{W} > 0$ at $p_W(0)$, so the wound decreases in size and the closure process starts, converges to a positive value, albeit smaller than $p_W(0)$(see Fig 1.b of suppl. section C);
    \item Regime III - the wound closes completely. This regime may be further divided into:
    \begin{itemize}
        \item Regime III$_a$ - $\partial \Tilde{\mathcal{F}}/\partial p_\mathrm{W} > 0$ at $p_W(0)$, which is above the bifurcation threshold, where the free energy has only one negative minimum and the wound closes, even for $\lambda_\mathrm{W} = 0$, being primarily driven by $p_\mathrm{0} > p^\text{bif}_\mathrm{0}$ (the tissue is in its fluid phase) (see Fig 1.c of suppl. section C);
        \item Regime III$_b$ - $\partial \Tilde{\mathcal{F}}/\partial p_\mathrm{W} > 0$ at $p_W(0)$, and there exists a positive minimum, which is above  $p_W(0)$. The wound closes successfully, as in regime III$_a$ (see Fig 1.d of suppl. section C). 
    \end{itemize}
\end{itemize}

The threshold separating regimes II and III$_b$ from regime III$_a$ corresponds to the bifurcation threshold, when the second and first derivatives of the function are null at the fixed point, and where the system undergoes a dynamical bifurcation. At linear order in $\varepsilon$, the threshold is given by Eq. (\ref{eqn:eq11}). The threshold between regime I and regimes II and III$_b$ corresponds to the closure threshold given by Eq. (\ref{eqn:eq10}).
To find the closure time, we first find a solution to Eq. (\ref{eqn:eq8}) near the stationary point $p_\mathrm{W} = 0$, using perturbative methods. The standard perturbation method gives secular terms which are hard to linearize, therefore we use the Lindstedt-Poincaré method \cite{nayfeh2011introduction}. The solution we find is of the form $p_\mathrm{W}(\tau) = p_\mathrm{W}^0 + \varepsilon p_\mathrm{W}^1$, where $\tau = t/\tau_\mathrm{p}$. The closure time is defined as the time for which $p_\mathrm{W}(\tau) = 0$. 
Using perturbative approaches it is possible to derive an analytic expression for the closure time that accounts for the size, given by:
\begin{equation}
\begin{split}
    \tau_\text{closure} &\approx -p^\text{crit}_\mathrm{0}\frac{ \log{\left(\frac{\lambda_\mathrm{W}}{\lambda_\mathrm{W}-\lambda^\text{clo}_\mathrm{W}}\right)}}{\gamma\left(p_\mathrm{0}-p^\text{crit}_\mathrm{0}\right)} + \mathcal{O}(\varepsilon), \text{ if } p_\mathrm{0} \neq p^\text{crit}_\mathrm{0} \\
    \tau_\text{closure} &\approx \frac{p^\text{crit}_\mathrm{0}}{\lambda_\mathrm{W}}\sqrt{\frac{N_\mathrm{W} \rho_\mathrm{W}}{\rho}} + \mathcal{O}(\varepsilon), \text{ if } p_\mathrm{0} \approx p^\text{crit}_\mathrm{0}.
    \label{eqn:eq12}
\end{split}    
\end{equation}
In practical applications, the additional terms have minimal impact when $\varepsilon \to 0$. To compare the predicted closure times with experimental and numerical results, we can use the dimensional closure time, given by the following expression:
\begin{equation}
    t_\text{closure} = \tau_\text{closure}\tau_\mathrm{p}.
    \label{eqn:eq13}
\end{equation}
If we assume $\lambda_\mathrm{W} \gg \lambda^\text{clo}_\mathrm{W}$, we can Taylor expand the expression for numerical fits (see suppl. section D).

\section{Numerical simulations}
\label{sec:3}

The mean-field calculations provide useful insights on the impact of the purse-string mechanics on wound healing. 
However, they assume that all cells are identical, neglecting spatial heterogeneities. 
Cells near the wound will deform more than those far from it.
To evaluate the robustness of the results, we performed numerical simulations of the vertex model.  
A regular hexagonal network of 225 cells is generated, with fixed boundaries corresponding to a strong tension at the boundaries of the tissue. 
The preferred  cell area is defined to be equal to the initial cell area and $\gamma = \tau_\mu = 1.429$ in dimensionless units. 
To form a wound, one $(N_\mathrm{W} = 1)$ or ten $(N_\mathrm{W} = 10)$ cells are removed at the center of the tissue.
\color{black}
In the relaxation dynamics, cells change shape and neighbors. 
The latter changes are topological in nature, with the underlying structure of the network changing as a result.
There are other relevant topological changes that may occur such as cell deaths and divisions.
Here, we only implement cellular neighbor exchanges, which are known as  $T_1$ transitions. 
This is a discontinuous process in which the vertices on a given edge swap neighbors (see suppl. section E) \cite{hashimoto2018}.

\color{black}
For $N_\mathrm{W} = 1$, we first measure the time-dependency of the wound perimeter normalized to its initial value, with time in $\tau_\mu$ units. 
Depending on the model parameters, we observe wounds opening (Suppl. Video 1), closing partially (Suppl. Video 2) or completely (Suppl. Video 3). 
However, some wounds exhibit a slight increase in size, before closing again to a smaller size. 
This increase or recoil is not captured by the mean-field calculations \cite{carvalho2021}.
\color{black}
\begin{figure}
    \centering
    \includegraphics[width=0.5\textwidth]{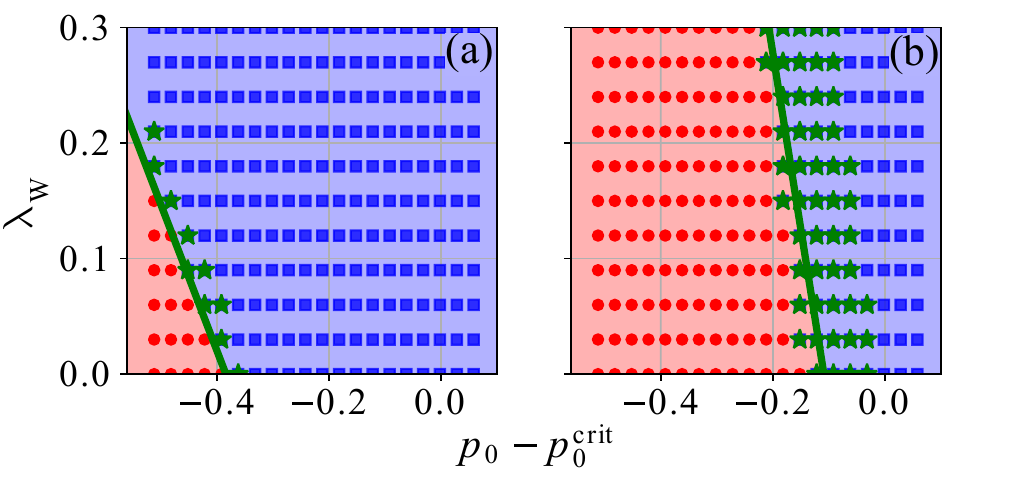}
    \caption{\footnotesize Parameter diagram from numerical simulations on an hexagonal network ($p_\mathrm{0}^\text{crit} = 3.722$), with red circles corresponding to openings, blue squares to closures and green stars to partial closures respectively. 
    Results shown for ($N_\mathrm{C}$ = 225 cells, $N_\mathrm{W}$ = 1 cell (a) and 10 cells (b)) with topological rearrangements.}
    \label{fig:fig4}
\end{figure}
\color{black} To identify the different regimes, for each pair of values of $p_\mathrm{0}$ and $\lambda_\mathrm{W}$, we assign a one if the wound closes or zero otherwise. 
We also distinguish between the cases where the wound closes partially or completely (Fig \ref{fig:fig4}.a). 
We find the closure threshold and fit the results using:

\begin{equation}
    \lambda^\text{clo}_\mathrm{W} = 1.265(3.338-p_\mathrm{0}).
    \label{eqn:eq14}
\end{equation} 
To compare with the mean-field predictions, we consider $\gamma = 1.429, \rho = \rho_\mathrm{W}, N_\mathrm{W} = 1$ and $p^\text{crit}_\mathrm{0} = 3.722$ in Eq. (\ref{eqn:eq10}), giving: 
\begin{equation}
    \lambda^\text{clo}_\mathrm{W} = 1.429(3.722-p_\mathrm{0}).
    \label{eqn:eq15}
\end{equation}
There is an 10.3\% shift on the threshold fit, and the resolution of numerical simulations, $\Delta p_\mathrm{0} = 0.03$, indicating a substantial shift in the estimated closure threshold. 
Partial closures occur near the closure threshold and correspond to closures with characteristic times much longer than the simulation time. This is contrary to our predictions for partial closures in regime II, where the stationary wound size is an open wound smaller than the initial size. We define \textit{long-time closures} as partial closures corresponding to the former, in contrast to \textit{true partial closures} which are predicted analytically to be in regime II. 
Given the resolution, for $N_\mathrm{W} = 1$ it is not possible to distinguish regime II from the other regimes.   

For $N_\mathrm{W} = 10$, the regime II is large enough to be observable with the considered resolution, and partial closures are observed relatively far from the closure threshold within the same interval range predicted by the analytical model (Fig \ref{fig:fig4}.b). The observation of partial closures due to long closure times is likely to be relevant in experimental settings. 
For $N_{\mathrm{W}} = 10$:
\begin{equation}
    \lambda^\text{clo}_\mathrm{W} = 3.05(3.613-p_\mathrm{0}).
    \label{eqn:eq16}
\end{equation} 
To compare with the mean-field predictions, we consider $\gamma = 1.429, \rho = \rho_\mathrm{W}, N_\mathrm{W} = 10$ and $p^\text{crit}_\mathrm{0} = 3.722$ in Eq. (\ref{eqn:eq10}), giving: 
\begin{equation}
    \lambda^\text{clo}_\mathrm{W} = 4.51(3.722-p_\mathrm{0}).
    \label{eqn:eq17}
\end{equation}
There is an 2.9\% shift on the threshold fit. There is still a large shift between the fitted and predicted results relative to the considered resolution, despite having a substantial decrease. The coefficients also have a large shift, but this comes from the assumption that $\rho_\mathrm{W} = \rho$ which is not true for $N_\mathrm{W} \neq 1$.

We do not observe sharp differences between regimes III$_a$ and III$_b$ since we only check if closure occurs.  
In the mean-field calculations, the separation between the two sub-regimes is given by the bifurcation threshold (Eq. (\ref{eqn:eq11})), but numerically we only observe the final outcome, which is successful closure in both regimes (see Fig 1.c and 1.d of suppl. section C). 
\color{black}

To account for spatial heterogeneities, we now perform simulations for irregular networks. 
$10$ networks of $225$ cells are generated (see suppl. section E), with fixed boundaries representing a strong tension at a distance. For each network, we define the preferred cell area $A_0$ as the average initial cell area. 
The average parameters across the networks are $\gamma = \tau_\mu  = 1.84$. 
Since $\varepsilon = 1/225$, we consider $p_\mathrm{0}^\text{bif} = p^\text{crit}_\mathrm{0} = 3.768$ and assume $\rho_\mathrm{W} = \rho$. 
As before, to form a wound, we remove one cell from the middle of the tissue, therefore $N_\mathrm{W} = 1$.

For $t \geq 2$, depending on the shape parameter $p_\mathrm{0}$ and effective gap tension $\lambda_\mathrm{W}$, the wounds either open (Suppl. Video 4), close partially (Suppl. Video 5) or completely (Suppl. Video 6), similar to what is observed for hexagonal networks. Recoil is observed for some closures in irregular networks, similar to what was observed for hexagonal networks. A key difference between the regular and irregular scenarios is the presence of a large variability of the simulation outcomes across samples. 

\color{black}

\begin{figure}
    \centering
    \includegraphics[width=0.475\textwidth]{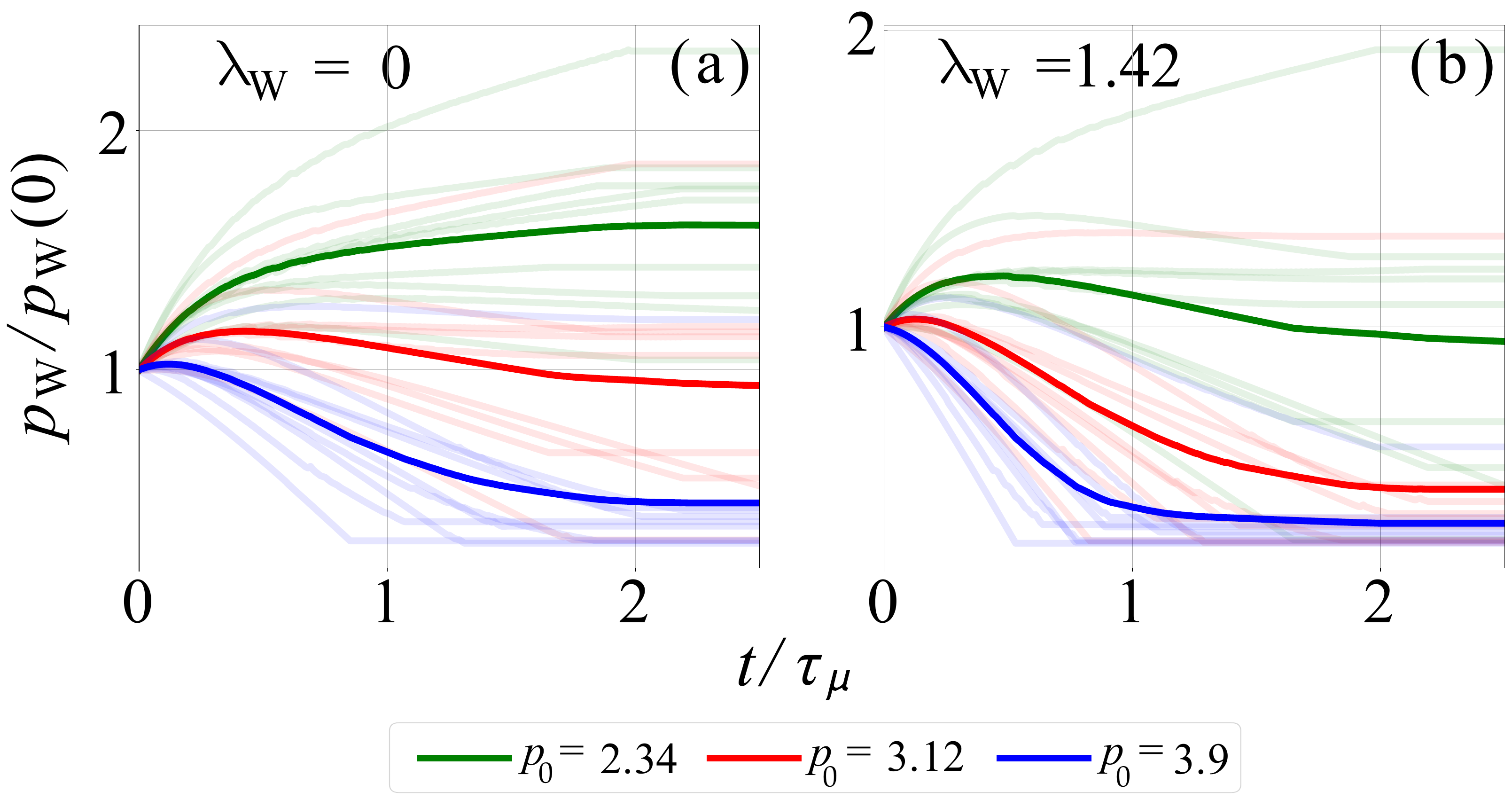}
    \caption{\footnotesize Time-dependence of the wound perimeter normalized by the initial perimeter, with time in $\tau_\mu$ units for a) $\lambda_\mathrm{W} = 0$ and  b) $\lambda_\mathrm{W} = 1.42$. The thicker curves are the averages across the generated networks, and the different colors correspond to different values of $p_0$. Results shown for irregular networks ($N_\mathrm{C}$ = 225 cells, $N_\mathrm{W}$ = 1 cell) without topological rearrangements.}
    \label{fig:fig5}
\end{figure}

\begin{figure}
    \centering
    \includegraphics[width=0.475\textwidth]{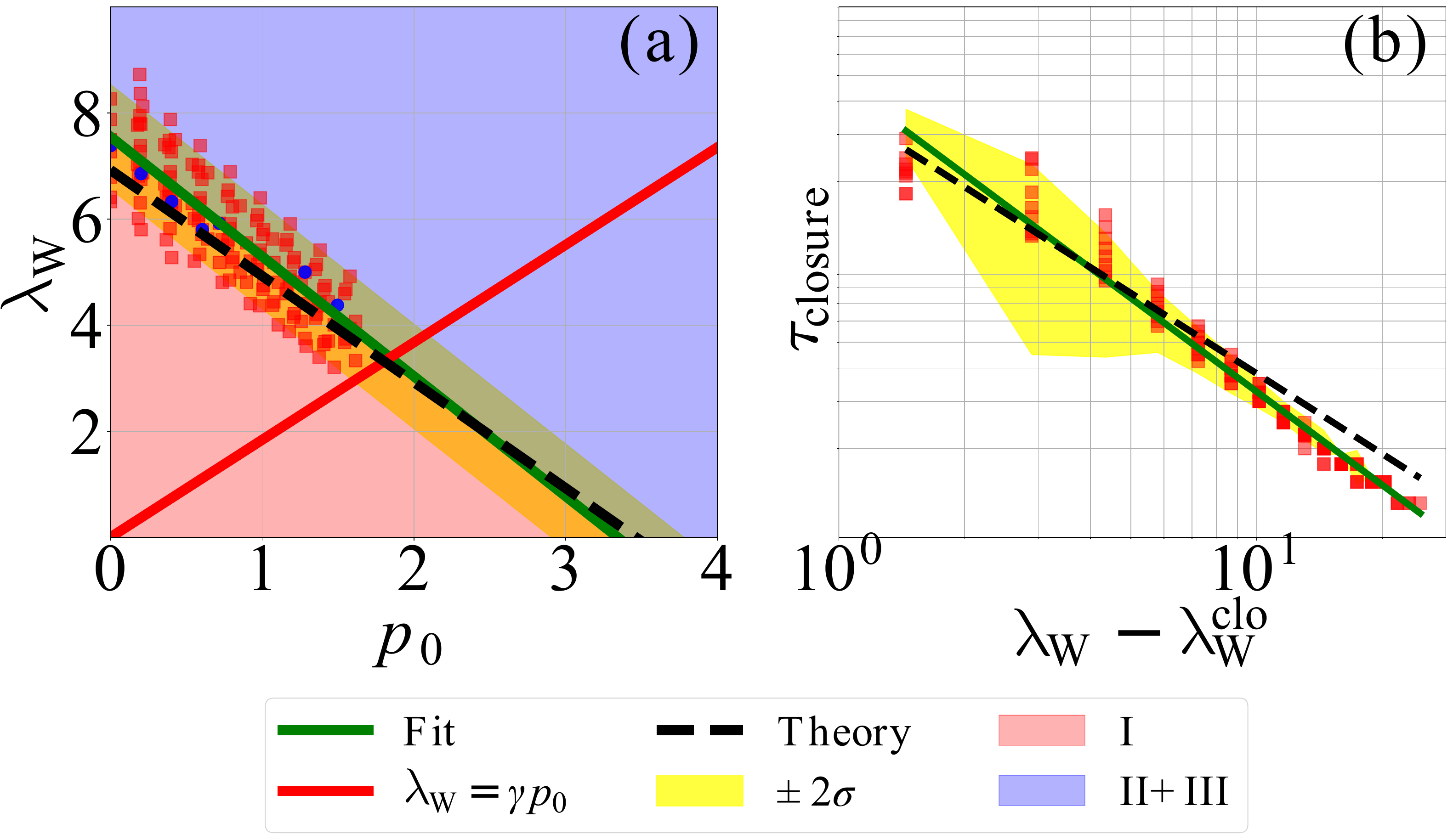}
    \caption{\footnotesize a) Closure threshold for different simulations (red squares), the numerical fit given by Eq.(\ref{eqn:eq18}) (green line) and the theoretical estimate given by Eq.(\ref{eqn:eq19}) (dashed black line). The blue circles correspond to values where partial closure occurred;  b) - Closure time (in log scale) of different simulations (red squares), the numerical fit given by Eq.(\ref{eqn:eq21}) (green line) and the theoretical estimate given by Eq.(\ref{eqn:eq22}) (dashed black line). Results shown for $N_\mathrm{c}$ = 225 cells and $N_\mathrm{W}$ = 1 cell with $T_1$ transitions. The yellow shaded area around the numerical fit corresponds to the confidence interval, defined as $\pm 2$ standard deviations.}
    \label{fig:fig6}
\end{figure}

We identify the different regimes, for each pair of values $p_\mathrm{0}$ and $\lambda_\mathrm{W}$, by assigning one if the wound closes or zero if not, as in the hexagonal case. 
We find the closure threshold (Fig. \ref{fig:fig6}.a, red squares), and fit the results using:
\begin{equation}
    \lambda^\text{clo}_\mathrm{W} = 2.26(3.345-p_\mathrm{0}).
    \label{eqn:eq18}
\end{equation}

To compare with the mean-field predictions, we consider $\gamma = 1.84, \rho = \rho_\mathrm{W}, N_\mathrm{W} = 1$ and $p^\text{crit}_\mathrm{0} = 3.768$ in Eq. (\ref{eqn:eq10}), giving: 
\begin{equation}
    \lambda^\text{clo}_\mathrm{W} = 1.84(3.768-p_\mathrm{0}).
    \label{eqn:eq19}
\end{equation}
There is a 10\% shift in the closure threshold (Fig.\ref{fig:fig6}.a, green and dashed black lines). Note that the resolution of the numerical simulations $\Delta p_\mathrm{0} = 0.39$ is of the same order as the difference between coefficients. We perform simulations without $T_1$ transitions as well, but the results are, within the error bars, the same (see Fig.3.a of suppl. section F). 

We cannot distinguish regimes II from III, despite observing wounds that only close partially (Fig.\ref{fig:fig6}, blue circles). This is in accordance with the results obtained for the hexagonal network with $N_\mathrm{W} = 1$. The presence of recoil in the wound perimeter may depend on whether the wound is in regimes III$_a$ or III$_b$ (see Fig 4 of suppl. section F). 

Based on the literature, we only consider non-negative values of purse-string tension for the simulations, corresponding to $\lambda_\mathrm{W} \geq \gamma p_\mathrm{0}$ (see suppl. section A) (see Fig.\ref{fig:fig6}.a, red line). 
We now apply our mean-field calculations to previous results (\cite{tetley2019}), for $\gamma = 0.04$ and $p^\text{crit}_\mathrm{0} = \rho^{-1/2}=3.81$. 
The authors consider a parameter space defined by the purse-string tension $\lambda^\text{ps}_\mathrm{W}$ relative to the adhesive tension $\lambda^\text{ad} = -2\gamma p_0$. We use $\lambda^\text{ad}$ as a stand-in variable for the adhesive tension.
We estimate a closure threshold of:
\begin{equation}
    \lambda^\text{ps}_\mathrm{W} = \lambda^\text{ad} + 0.15.
    \label{eqn:eq20}
\end{equation}
Their numerical results correspond to a slope of 1 with an intercept of $-0.12$ in the adhesive tension, showing similar behavior to our analytic estimate.  

Finally, we compute the closure time for the different simulations.
We fit the results with the following expression:
\begin{equation}
    t_\text{closure} = 0.766 (\lambda_\mathrm{W}-\lambda^\text{clo}_\mathrm{W}) ^{-1.05}.
    \label{eqn:eq21}
\end{equation}
The closure time estimated from Eq. (\ref{eqn:eq12}) is, after rescaling, of the same order of magnitude as fitted results in Eq.(\ref{eqn:eq21}) (Fig.\ref{fig:fig6}.b). If we assume $\lambda_\mathrm{W} \gg \lambda^\text{clo}_\mathrm{W}$, and $\tau_\mathrm{p} = 0.19$, we obtain:

\begin{equation}
    t_\text{closure} = 0.715 (\lambda_\mathrm{W}-\lambda^\text{clo}_\mathrm{W}) ^{-1},
    \label{eqn:eq22}
\end{equation}
which is in agreement with the numerical fit (see suppl. section D).
The exclusion of $T_1$ rearrangements does not impact the closure time significantly (see Fig.3.b of suppl. section F). 

\section{Conclusion}
\label{sec:4}
Our study delves into the dynamics of wound healing in biological tissues, shedding light on the interplay between cellular properties and closure mechanisms.
Through analytical mean-field study, we provide insight into the underlying dynamics.
We find that in the absence of an effective wound tension $(\lambda_\mathrm{W} = 0)$, the shape parameter $p_\mathrm{0}$ is the primary determinant of wound closure. 
The closure only occurs if $p_\mathrm{0}$  is below a bifurcation threshold $p_\mathrm{0}^\text{bif}$, which in the limit $\varepsilon \rightarrow 0$, coincides with fluid transition of the tissue, as reported previously \cite{noppe2015, tetley2019}.
In the presence of an effective wound tension $(\lambda_\mathrm{W} \neq 0)$, the closure process changes by reducing the effective $p_\mathrm{0}^\text{bif}$. For some initial perimeter $p_\mathrm{W}(0)$ and values of $p_\mathrm{0}$ below $p_\mathrm{0}^\text{bif}$, wound closure is only possible if $\lambda_\mathrm{W}$ exceeds a minimum value $\lambda_\mathrm{W}^\text{clo}$, which is independent of the tissue size, as derived exactly. 
The dependence of the closure threshold on the initial wound size is consistent with prior research, where for a fixed wound tension, it is reported a critical wound size for successful healing \cite{vedula2015, noppe2015}.

The bifurcation and closure thresholds in the space of $p_\mathrm{0}$ and $\lambda_\mathrm{W}$ divide it into \textit{three regimes}. 
Regime {I} corresponds to situations where closure is not possible due to a wound size exceeding the closure threshold. Regime {II} describes \textit{partial closure}, indicating that the wound decreases in size but is unable to close completely. Regime {III} indicates \textit{successful closure}.
The latter is further divided into regime III$_a$, where the closure occurs beyond the bifurcation threshold, and regime III$_b$, which corresponds to a closure that occurs prior to the bifurcation threshold. 
The boundary between regimes  I and  {II} is given by the closure boundary.  
Through mean-field calculations, we derived an analytic expression for closure time.
It diverges near the threshold defined by $\lambda^\text{clo}_\mathrm{W}$, indicating longer closure times for processes occurring close to it. 
Moreover, when $(p_\mathrm{0}, \lambda_\mathrm{W})$ falls below this threshold, closure times become undefined, indicating a failure in the closure process. 

Numerical simulations corroborate the mean-field results, albeit with significant variance of the numerical results. 
The closure threshold persists even in systems with finite size, showing its robustness in the vertex model, which we verify numerically. The presence or absence of T$_1$ rearrangements does not significantly alter the results. For the hexagonal network, we observe the different regimes, with regime I and III visible for all cases, and regime II only visible for $N_\mathrm{W} = 10$. There is a significant shift between the predicted and observed closure thresholds, being more significant for smaller wounds.  Using the parameters provided by previous works, we manage to derive the closure boundary consistent to the one obtained numerically \cite{tetley2019}.
The variability across multiple networks indicates that rather than a sharp transition between the two regimes, a smooth, fuzzy boundary may be a better representation of closure process.
 
We observed an initial recoil before closure, a phenomenon we could not predict from the mean-field results. 
We were not able to observe a significant distinction between regimes III$_a$ and {III}$_b$. 
We observed partial closures near regime II in the parameter space, but we also observed partial closures near the closure threshold but far from regime II. These partial closures, which we define as \textit{long-time closures}, are qualitatively distinct from the analytical \textit{true partial closures}. However, long-time closures are expected to be very relevant in experimental settings.

From the mean-field approach, we managed to establish the conditions under which the purse-string mechanism alone is sufficient for wound closure and whether it is possible in its absence. 
Previous experimental results \cite{carvalho2018} showed the link between mutations in occluding junctions affecting actin-myosin purse polymerization. 
Our results allow us to frame those results as an interplay between $p_0$ and $\lambda_W$. 
For example, if for a wild-type case, the effective gap tension is slightly above the closure threshold, a reduction induced by mutation is sufficient to cause closure failure. 
If the effective wound tension of the mutant lies below the closure threshold, not only closure fails, but the wound increases in size to possibly drastic extents. 

These results show that a mean-field approach, while simple, manages to capture several significant features of the wound healing process, and to derive analytical expressions that can be empirically tested, with parameters that are physically interpretable and biologically relevant. Our numerical results also show the importance of spatial heterogeneities, neglected by the mean-field model, in particular when determining the closure threshold in parameter space, or in explaining wound recoil.
These findings open the door to further explorations of the dynamics of tissue closure processes, shedding light on the complex interactions that govern these crucial biological events.

\section{Acknowledgements}
\begin{flushleft}
The authors acknowledge financial support from the Portuguese Foundation for Sciences and Technology (FCT) under Contracts no. UIDB/00618/2020 (https://doi.org/10.54499/\href{https://doi.org/10.54499/UIDB/00618/2020}{UIDB/00618/2020}), UIDP/00618/2020
(https://doi.org/10.54499/ \href{https://doi.org/10.54499/UIDP/00618/2020}{UIDP/00618/2020}) and UIBD/154422/2022 (https://doi.org/10.54499/\href{https://doi.org/10.54499/UIBD/154422/2022}{UIBD/154422/2022}).
\end{flushleft}

\bibliography{biblio}

\end{document}